\documentclass[aps,prl,fleqn,twocolumn,longbibliography,footinbib]{revtex4-2}

\usepackage{graphicx}
\usepackage{amsmath} 
\usepackage{color} 
\definecolor{gray}{rgb}{0.5,0.5,0.5}
\definecolor{darkblue}{rgb}{0,0,0.5}
\definecolor{darkred}{rgb}{0.5,0,0}
\definecolor{darkgreen}{rgb}{0,0.5,0}

\def\i{\mathrm{i}}
\def\d{{\rm d}}

\def\e#1{{\rm e}^{#1}}

\def\vec#1{\boldsymbol{#1}}
\def\vK{\vec{K}}
\def\vA{\vec{A}}
\def\vF{\vec{F}}
\def\ve{\vec{e}}
\def\vq{\vec{q}}
\def\vqi{\vec{q}}
\def\qi{q}
\def\vqt{\widetilde{\vec{q}}}
\def\vQt{\widetilde{\vec{Q}}}
\def\qt{\widetilde{q}}
\def\Pc{P}

\def\Hlin{\overline{H}}
\def\Elin{\overline{E}}
\def\Vlin{\overline{V}}
\def\vAlin{\overline{\vA}}

\def\tint{\mbox{\large$\int$}}
\def\bigs#1{\scalebox{1.1 }{\ensuremath#1}}

\begin{document}

\title{Intense-pulse dynamics of massless Dirac electrons}
\author{Hamed Koochaki Kelardeh}
\author{Ulf Saalmann} 
\author{Jan M. Rost}
\affiliation{Max-Planck-Institut f{\"u}r Physik komplexer Systeme,
 N{\"o}thnitzer Str.\ 38, 01187 Dresden, Germany}
\date{\today}

\begin{abstract}\noindent
We identify and describe how intense short light pulses couple to massless Dirac fermions in two-dimensional systems.
The ensuing excitation dynamics exhibits unusual scaling with the wavelength of the light due the linear dispersion of the band structure and the fact that light coupling is efficient only close to the Dirac points. 
We exploit these features to achieve valley polarization of more than 70\,\% with simple pulse shapes.
Quantitative results are given for pristine graphene.
\end{abstract}

\maketitle

\noindent
Electrons in two-dimensional (2D) materials carry pseudospins (of opposite sign) at the minima $\vK$ and $\vK'$ (valleys) of the valence band in the 1st Brillouin zone \cite{ca13,le10}. The long lifetime of these degrees of freedom has given their manipulation and transport, in short valleytronics, a boost in the quest to search for usable quantum information encoding \cite{hoqi13,scyu+16,vine+18}.
Most work has been devoted to finite-mass Fermions in gapped materials \cite{mcko13,lasc+18,jisi+21,kooc+21,wawe+21} or gapless material under static fields \cite{wali+17,frdi21} 
since they render addressing $\vK$ and $\vK'$ points separately by external driving easier, while breaking the equivalence of the two Dirac points in gapless systems by optical pulses seemed to be impossible \cite{yaxi+08,leso+15,mone+19}, although rich dynamical behavior in graphene induced by strong short pulses has been predicted \cite{keap+15}.
Indeed, a carefully chosen pulse form, with two colors and time-dependent polarization, adapted to the hexagonal geometry of the band structure and maintaining an electric field component during the pulse to break the equivalence of $\vK$ and $\vK'$ \cite{mrji+21} has been demonstrated to achieve valley polarization in pristine graphene recently. 

In the following, we will uncover the general mechanism of coupling Dirac dynamics to
intense laser pulses which is ruled by (i) the vector potential $A_0$ of the light pulse in units of the distance $\Delta$
between the Dirac points $\vec\Delta = \vK-\vK'$ in momentum space and (ii) by time and momentum scaled with the square root of the light frequency (or wavelength), $\tau = \sqrt{\omega}t$ and $Q =q/\sqrt{\omega}$. 
The scaling emerges from the fact that coupling to the light only occurs in the vicinity of the Dirac points with linearized dynamics and dipole matrix elements about these points exhibiting this scaling. Furthermore, pulses with $A_0\approx \Delta$ are most efficient for inducing large valley polarization since the latter relies on transporting excitation from valley domain $\vK$ to $\vK'$ or vice versa where the domain separation in momentum space is characterized quantitatively by $\Delta$. 

As an application of this general mechanism we will demonstrate that a conventional half-cycle pulse with linear polarization along $\vec\Delta$ can achieve
substantial valley polarization (VP) while including a weaker pre-pulse (``pedestal'') and thereby making use of the subtleties of the mechanism can lead to almost 100\,\% VP.

To be specific we will work with pristine graphene in the usual tight-binding two-band description \cite{le10} (conduction and valence band), coupled in dipole approximation to a classical electromagnetic field with the time-dependent vector potential
\begin{equation}
 \label{eq:lightpulse}
 \vA(t) = \vA_0 \e{-2\ln2\, t^2 /T^2}\cos(\omega t) \,.
\end{equation}
Mostly, we will apply sub-cycle pulses with a duration of $T{=}T_\omega/5$, where $T_\omega = 2\pi/\omega$ is the laser period.

We solve the time-dependent Schr\"odinger equation for the time-dependent two-band Hamilton operator
\begin{subequations}
\begin{align}\label{eq:hamil}
\end{align}
\end{subequations}
whereby the C--C distance is $a\,{=}\,1.42$\,\AA\ \cite{le10}.
Inclusion of relaxation is easily possible by solving the Liouville–von-Neumann equation for the single-particle density matrix. However, this is not necessary in the present context due to the short light pulses.

\begin{figure}[t!]
\centering
\includegraphics[width=0.9\columnwidth]{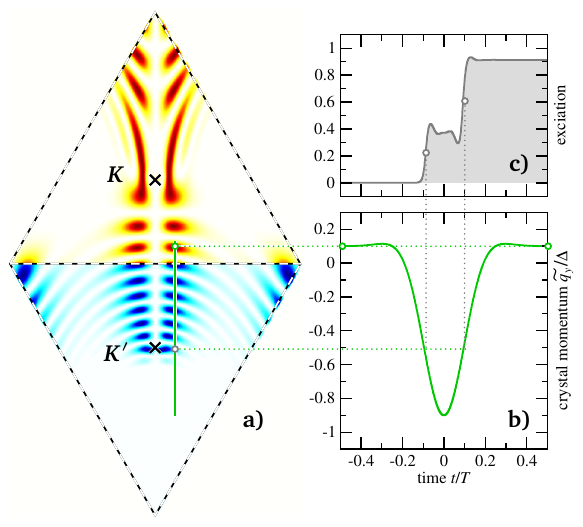}
\caption{\textbf{a)} Excitation probability $\rho(\vec q)$ as a function of the initial crystal momentum $\vqi$ for
the half-cycle pulse of \eqref{eq:lightpulse} with
$\lambda{=}2.4\,\mu$m and field strength $A_0{=}\Delta$ and a pulse length of $T{=}T_\omega/5$. The Dirac points in the centers of the triangular unit cells are marked by crosses. 
\textbf{b)} Time-dependent crystal momentum $\qt_{y}(t)=\qi_{y}+A(t)$ in vertical direction.
\textbf{c)} Corresponding excitation probability $|a_{\rm c}(t)|^2$ for the initial condition $\vqi=\Delta(1.0,0.1)^\mathrm{T}$. 
}
\label{fig:sketch}
\end{figure}%
We begin with the excitation induced by a half-cycle pulse, with $T{=} T_\omega/5$ in Eq.\,\eqref{eq:lightpulse}, linearly polarized parallel to $\vec\Delta=\Delta \ve_y$  as this dynamics reveals the ruling principles of intense pulse coupling to massless Dirac electrons quite clearly. Figure~\ref{fig:sketch}a resolves this excitation probability $\d^2\Pc/\d\qi_x\d\qi_y \equiv \rho(\vqi)$ with respect to all initial conditions in the Brillouin zone, discriminated (in blue and red) according to the $\vK$ and $\vK'$ domain (lower and upper triangle). One sees interference patterns along the polarization direction (but not on the $\vec\Delta$-line directly!) with denser fringes in the lower triangle indicating a larger phase accumulation of the driven electron wave than in the upper triangle. 
The very fact of the clean pattern suggests interference of two amplitudes with a well-defined phase difference. 
The underlying momentum space trajectory $\vqt(t) = \vqi + \vA(t)$ for a specific initial condition $\vqi = \Delta(1.0,0.1)^\mathrm{T}$ in Fig.\,\ref{fig:sketch}b reveals that indeed, there are two instances of closest proximity to the $\vK$-point (and hence two excitation ``bursts'') with $ \qt_y(t) \equiv 0.1\Delta + A_y(t) = K_y$, one in the rising and one in the falling part of the pulse (see lower horizontal green-dotted line).
At both time instants the excitation suddenly rises, see Fig.\,\ref{fig:sketch}c.
Clearly, it is crucial to come close to the Dirac point for excitation. This also explains why there are no excitations for initial conditions with $\qi_y < K'_y$ since the corresponding laser-driven trajectories do not visit any Dirac point.
Moreover, the phase difference between the two bursts for this initial condition is close to $5\pi$ which effects constructive interference. 
Since the transition energy along the trajectory is larger below the $\vK'$ point (compared to the one between $\vK$ and $\vK'$ point)  the  phase accumulated between the two excitations is larger there. Consequently,  trajectories starting above $\vK'$ yield denser fringes than those staring above $\vK$, as can be seen in Fig.\,\ref{fig:sketch}a.

\begin{figure}[b!]
 \centering
 \includegraphics[width=\columnwidth]{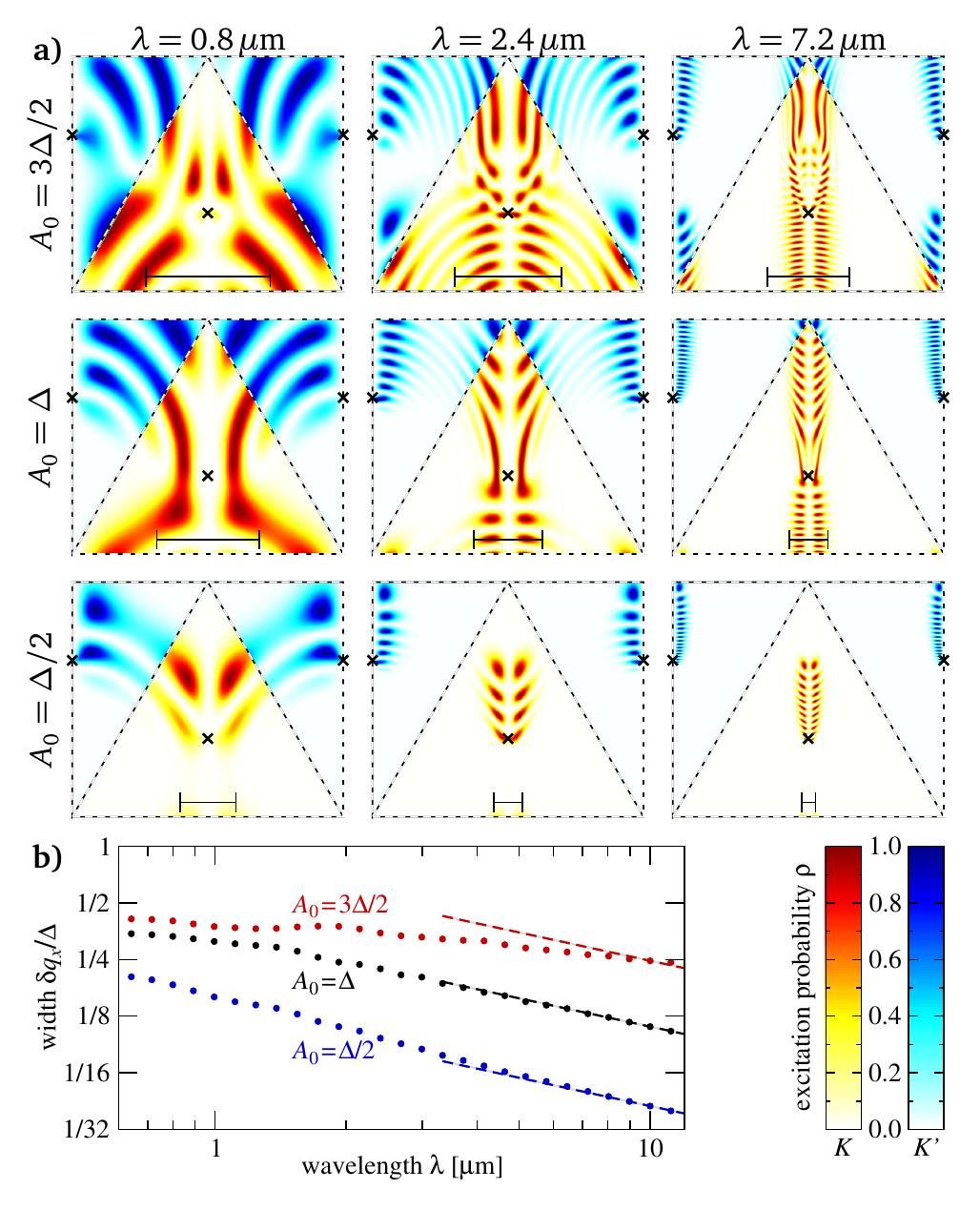}
 \caption{{\bfseries a)} Excitation probability $\rho$ as a function of the initial crystal momentum $\vq$ for the pulse \eqref{eq:lightpulse} with a pulse length of $T{=}T_\omega/5$ and
 for various wavelengths $\lambda$ and field strength $A_0$. The Dirac points in the centers of the triangular unit cells are marked by crosses. (For a compact representation we have, in contrast to Fig.\,\ref{fig:sketch}, nested the unit cells.)
 The width $\delta q_x$ of the excitation pattern is marked with a black line at the bottom of each contour plot. It is defined as second moment $\delta q_x\equiv \sqrt{m_2/m_0}$ with $m_k\equiv\int_{\triangle\vK}\d\qi^2 \,\qi_x{\!}^k\,\rho(\vqi)$.
 {\bfseries b)} Pattern width $\delta q_x$ as function of the wavelength $\lambda$ for three values of $A_0$. 
 The dashed lines show $\delta\qi_x\,{=}\,\alpha/\sqrt{\lambda}$, with $\alpha$ fixed at $\lambda{=}10\,\mu$m. The color scale at the bottom right is the same as in Figs.\,\ref{fig:sketch}a and \ref{fig:pedestal}a.
 }
 \label{fig:lin-t02-comp}
\end{figure}%

Having established that intense pulse optical excitation of massless Dirac electrons occurs only close to the Dirac points, we next turn to the scaling properties of the excitation dynamics. 
To this end we consider excitation with the same pulse shape as before but at different wavelengths and for three different $A_0$ of the light for all initial conditions in the first Brillouin zone, shown in Fig.\,\ref{fig:lin-t02-comp} with a compactified representation. Note that $\vK'$ points are now located at the left and right boundaries of the rectangle in the upper half (marked with little crosses in the figure). One sees that the structures have the same shape but are ``zoomed out'' from left to right for longer wavelengths, most obvious for the two lower rows while in the first row additional interference due to the large excursion of the trajectories masks the underlying similarity of the pattern. 

Excitation happening only close to the Dirac points suggests to linearize the dynamics in momentum space and time close to a Dirac point by Taylor expanding about it, $\vq_{K}\,{\equiv}\, \vq - \vK$ with $\vK\,{=}\, [2\pi/3a](1,{+}1/\sqrt{3})^\mathrm{T}$ or $\vq_{K'}$ with $\vK' \,{=}\, [2\pi/3a](1,{-}1/\sqrt{3})^\mathrm{T}$. 
Hence $H_{\vq}$ from \eqref{eq:hamil} may be approximated close to $\vK$ by the familiar relativistic Hamiltonian for massless spin-1/2 particles $\Hlin_{K}\,{=}\, v_{\rm F}\,\vec{\sigma}\,{\cdot}\,\vq_{K}$
\footnote{Indeed there are three linearized Hamilton operators for each $\vK$ and $\vK'$, which differ by a phase factor in front of the non-diagonal elements of $\Hlin$. Those phase factors are irrelevant for both the eigenenergies and the Berry-connection matrix.} in terms of the Pauli matrices $\vec{\sigma}\,{=}\,(\sigma_x,\sigma_y)^\mathrm{T}$ and the speed of light replaced by the Fermi velocity $v_{\rm F}\,{=}\, 3g_0 a/2$. 
The eigenvalues of $\Hlin_{K}$ are $\Elin_{\rm c,v}(\vq_{K})\,{=}\,\pm v_{\rm F}|\vq_{K}|$ with eigenvectors, often referred to as Houston basis \cite{ho40}, $\Vlin_{\rm v,c}(\vq_K) \,{=}\, (\pm[q_{Kx}{+}\i q_{Ky}] /|\vq_K|,1)^\mathrm{T}/\sqrt{2}$ for valence (v) and conduction (c) band, respectively. 

Within this approximation the time-dependent dipole-coupled Hamilton operator in the Houston basis can be expressed as
\begin{equation}\label{eq:hlin}
 \Hlin(\vqt,t) = v_{\rm F}|\vqt_K(t)| \sigma_z
+\frac{\vF(t)\cdot[\ve_z{\times}\,\vqt_K(t)]}{2|\vqt_K(t)|^2}\sigma_x\,,
\end{equation}
where the second term in $\Hlin$ is a real, reduced Berry-connection matrix 
\footnote{The full Berry-connection matrix in the Houston basis is given by $\vAlin_{jk}\equiv\Vlin^*_j\,\i\frac{\d}{\d\vq}\Vlin_k$ with $j,k={\rm v,c}$.}, from which the diagonal elements have been omitted as they are the same and shift
only the total energy of $\Hlin$, while the off-diagonal terms couple valence and conduction band through the electric field $\vF(t)=\frac{\d}{\d t}\vA(t)$ of the light.

From Eq.\,\eqref{eq:hlin} one sees that only close to the $K$-point, where $|\vqt_K(t)|$ is small, transitions to the other band happen through the dipole coupling. The corresponding time-dependent dynamics is governed by the approximate Hamilton operator $\Hlin$ valid close to Dirac points with intriguing scaling properties regarding the dependence on the light frequency $\omega$: Consider a trajectory in the vicinity of $\vK$, e.\,g., $\vqt_K(t)=(b_x,c_y \omega t)^\mathrm{T}$, which passes $\vK$ at $t{=}0$ at a distance $b_x$ and a velocity $c_y\omega$. This trajectory corresponds to
a linearly polarized pulse along $\ve_y$ with frequency $\omega$ as in Fig.\,\ref{fig:lin-t02-comp}, but the following argument holds for any close encounter of a Dirac point.
In terms of a scaled time $\tau$ and a scaled momentum $B_x$
\begin{equation}\label{eq:scales}
 \tau=\sqrt{\omega}t\quad\mbox{and}\quad B_x=b_x/\sqrt{\omega}\,,
\end{equation}
 we get from \eqref{eq:hlin}
\begin{equation} \label{eq:hlinscal}
\Hlin(\vqt_K,t) = \Hlin(\sqrt{\omega}\,\vQt_K,\tau/\sqrt{\omega}) = \sqrt{\omega}\,\Hlin(\vQt_K,\tau)
\end{equation}
with $\vQt_K(\tau){=}(B_x,c_y\tau)^\mathrm{T}$. 
Obviously, \eqref{eq:hlinscal} gives rise to a time-dependent Schr\"odinger equation $\big[\Hlin(\vQt_K,\tau)-\i\frac{\partial}{\partial\!\tau}\big] \psi(\tau)=0$ which is invariant against changes of the frequency $\omega$. It
directly explains the scaling of the widths $\delta q_x$ of the excitation pattern for different wavelengths, respectively frequencies, encountered in Fig.\,\ref{fig:lin-t02-comp}. This is a universal result for intense light, dipole-coupled to massless electrons.

As a first application of this result we can formulate simple laser-pulse shapes to control valley polarization with high efficiency. The latter can be quantified as $\eta$ by the relative excitation probability difference of initial conditions in the $\vK$ (red in Figs.\,\ref{fig:sketch} and \ref{fig:lin-t02-comp}) and $\vK'$ domain (blue in Figs.\,\ref{fig:sketch} and \ref{fig:lin-t02-comp}),
\begin{subequations}\label{eq:VP}\begin{align}
\eta &\equiv (\Pc_{K} - \Pc_{K'})/(\Pc_{K} + \Pc_{K'})\,,
\intertext{with}
\label{Eq:NkNKprime}
\Pc_{K} & = \tint_{\!\!\!\triangle\vK}\,\d\qi^2 \rho(\vqi),\quad
\Pc_{K'} = \tint_{\!\!\!\triangle\vK'}\,\d\qi^2 \rho(\vqi)\,
\end{align}\end{subequations}
integrated over the triangular $\vqi$-domains $\triangle\vK$ and $\triangle\vK'$ around $\vK$ and $\vK'$, respectively.
Note that our definition of $\eta$ allows values between $-1$ and $+1$ for polarization of the $\vK$ or $\vK'$ pseudospin, respectively, while in other definitions
\cite{mrji+21} the values of $\eta$ range from $-2$ to $+2$. 

With the half-cycle pulse discussed so far, one can see already from Fig.\,\ref{fig:lin-t02-comp} that due to the $\sqrt\omega$ scaling smaller frequency (larger wavelength) confines the excitation tighter to the electron excursion $\vqt(t)$ driven by the light pulse and therefore promises a higher degree of control. 
Combined with a laser-driven excursion compatible with the geometry of the Brillouin zone and particular its scale given by the separation $\Delta$ of the Dirac points, one can achieve a large asymmetry in the excitation of the $\vK$ versus the $\vK'$ domain, manifest in the dominance of red pattern in comparison to blue ones, most prominent for the right panel in the middle row of Fig.\,\ref{fig:lin-t02-comp} which corresponds to the largest wavelength displayed and the expected optimum near $A_0=\Delta$. 
One sees that for $A_0<\Delta$ the initial conditions in the $\vK$ as well as in the $\vK'$ domain are not exhausted for excitation (lower right panel) while for $A_0>\Delta$ also initial conditions from the $\vK'$ domain get excited through the trajectory passing by $\vK$ (blue intensity in the middle on the top of the right upper panel) lowering the contrast of valley polarization.
The latter is in general true for shorter wavelength (left row in Fig.\,\ref{fig:lin-t02-comp}), where initial conditions from the $\vK$ and $\vK'$ domain are almost equally excited in the first Brillouin zone. 

Next to the ``resonance'' condition $A_0\approx \Delta$ and long wavelength, the half-cycle nature of the pulse, containing amplitude dominantly only in one direction, is of course crucial since it breaks the equivalence of $\vK$ and $\vK'$ points for the laser-driven dynamics. 
Longer pulses with amplitudes in both directions restore this equivalence and diminish valley polarization. 
Not necessary, however, is a helicity character of the pulse (circular polarization, e.g.) which was thought to enhance or decrease the interaction with the pseudospin degrees of freedom depending on the direction of rotation. 

\begin{figure}[h!]
 \centering
 \includegraphics[width=0.75\columnwidth]{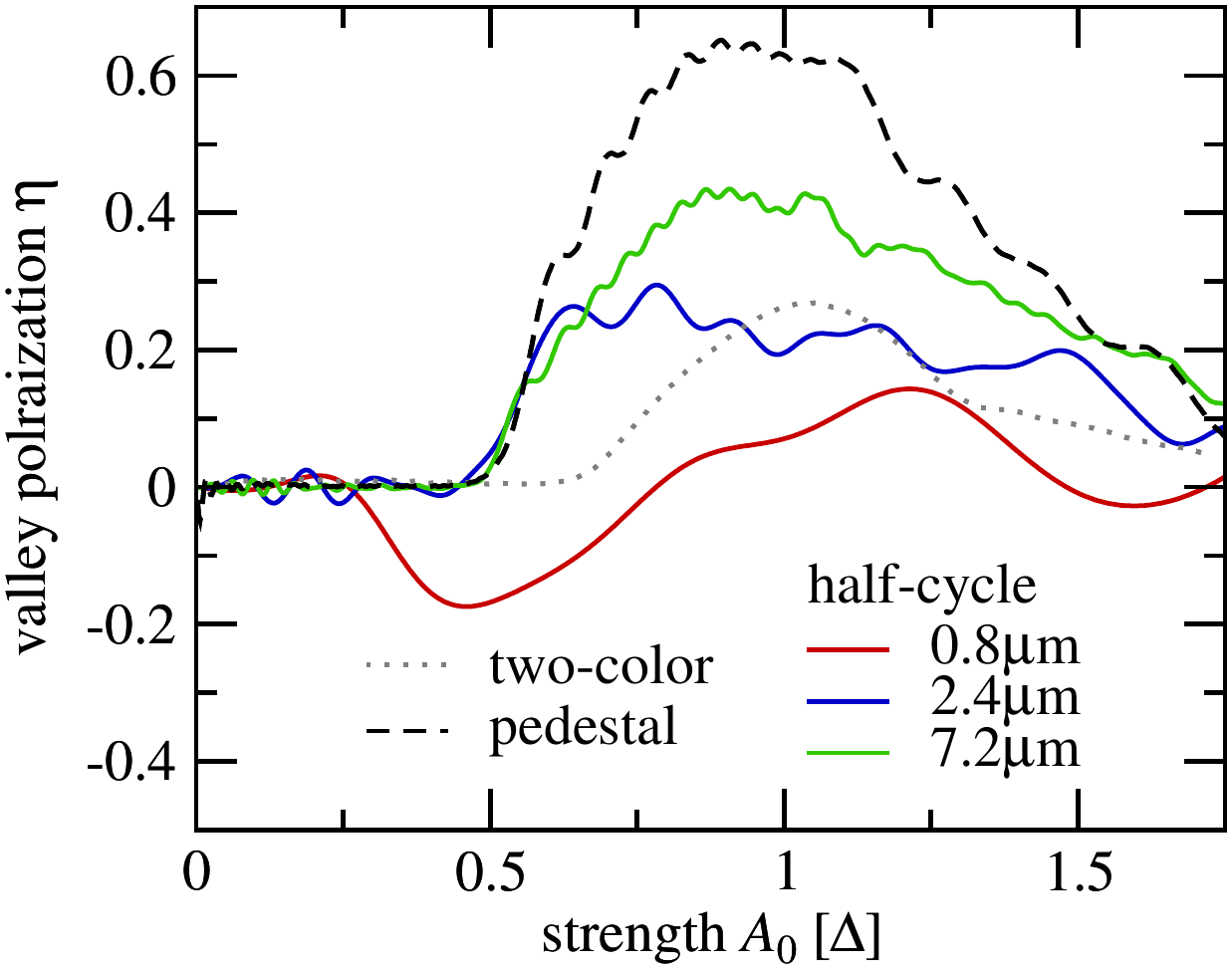}
 \caption{Valley polarization for half-cycle pulses with 3 wavelengths as a function of the peak vector potential in unit of the $\vK{-}\vK'$ point distance $\Delta$. For comparison we show results for a two-color pulse \cite{two-color} and a half-cycle pulse with a pedestal pulse according to Eq.\,\eqref{eq:ped-pulse} with $\xi\,{=}\,8$, $T\,{=}\,T_{\omega}/5$ and $\lambda\,{=}\,2.4\,\mu$m.
 }
 \label{fig:vp}
\end{figure}%

The simple half-cycle pulse discussed so far is also more efficient at a wavelength of $\lambda\,{=}\,7.2\,\mu$m with $\eta\,{=}\,43\,\%$, than a two-color clover-shaped pulse \cite{mrji+21} with $\eta\,{=}\,27\,\%$, as can be seen in Fig.\,\ref{fig:vp}, where VP is shown as a function of $A_0$ for different pulse forms and different wavelengths for the half-cycle pulses. Apart from the shortest wavelength (800\,nm) where the blurred fringe pattern leads to excitation delocalized in $\vqi$-space and therefore to irregular changes of $\eta$ with increasing $A_0$, all other pulses behave qualitatively similar with the achieved VP as a function of $A_0$: VP sets in for $A_0\,{>}\,\Delta/2$ and reaches a maximum around $A_0\,{=}\,\Delta$. The delayed onset becomes understandable by realizing that a Dirac point is located at a distance larger than $\Delta/2$ from any edge of its triangular domain. However, only if initial conditions $\vqi$ from a \emph{different} Dirac domain reach the $\vK$ point through a laser-driven trajectory which crosses the boundary of the triangular domain of $\vK$, one can expect finite VP.

The half-cycle pulse
\begin{equation}\label{eq:ped-pulse}
 A_\xi(t)=\frac{A_0}{2}\big[\e{-2\ln2\, t^2 /T^2}+\e{-2\ln2\, t^2 /\xi^2 T^2}\big]. 
\end{equation}
with a pedestal, another half-cycle pulse that is $\xi$ times longer, however, stands out as it produces a VP of more than 70\,\% (dashed line in Fig.\,\ref{fig:vp} with $\xi = 8$) which may  reach  close to 100\,\% by optimizing the pulse shape which we have not done since here we are interested in the principles of controlling VP with intense pulses. 

\begin{figure}[h!]
 \centering
 \includegraphics[width=\columnwidth]{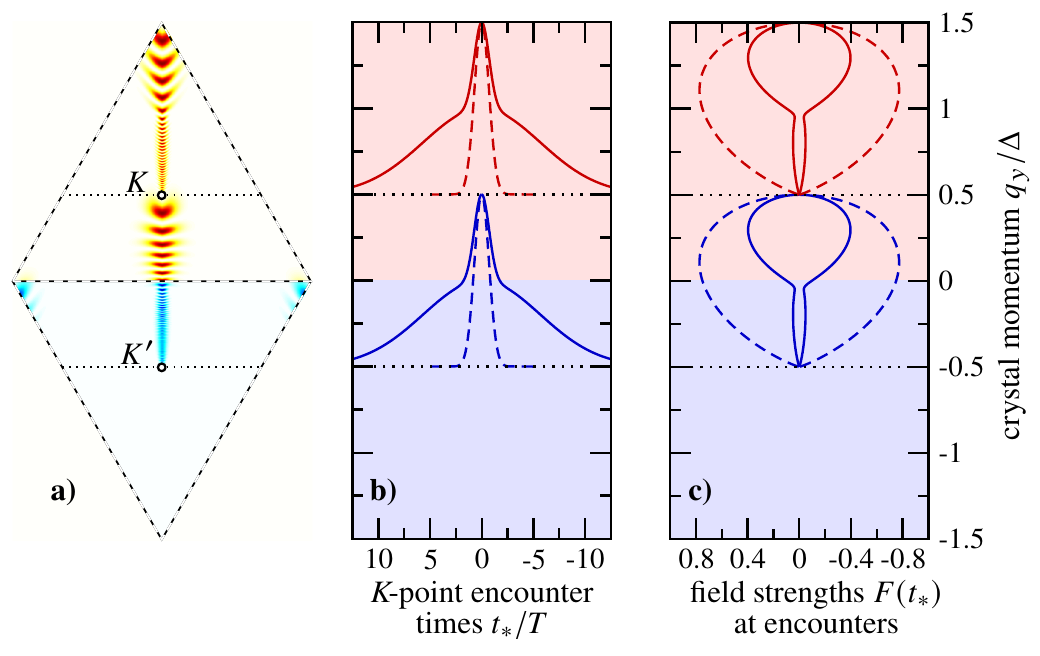}
 \caption{\textbf{a)} Excitation probability as a function of the initial crystal momentum $\qi_y$ for the pedestal pulse from Fig.\,\ref{fig:vp} for $A_0=\Delta$.
 \textbf{b)} The two times $t_{*}$ (rising and falling part of the pulse) when the ``trajectory'' $\vqt(t)$ encounters $\vK$ (red) or $\vK'$ (blue), respectively. The solid lines show the times for the pulse \eqref{eq:ped-pulse}, the dashed ones for the corresponding half-cycle pulse without pedestal. Note that for $\qi_{y}=-\frac{3}{2}\ldots-\frac{1}{2}\Delta$ the trajectory never ``hits'' any $\vK$ or $\vK'$ point.
 \textbf{c)} Field strengths $F$ at those encounter times; large $|F(t_{*})|$ imply large excitation probabilities.
 }
 \label{fig:pedestal}
\end{figure}

In that respect, it is important to fulfill one more criterion for efficient excitation when passing a Dirac point at time $t^*$: The field strength $F(t^*)$, cf.\ Eq.\,\eqref{eq:hlin}, must be large, otherwise even a large dipole matrix element does not help. 
Vice versa, large field strengths can compensate to some extent small dipole matrix elements giving rise to ``uncontrolled'' excitation, further away from the Dirac points. The pulse \eqref{eq:ped-pulse} fulfills this criterion as one can see from Figs.\,\ref{fig:pedestal}b and c. There, the initial conditions of the $\vK$ and $\vK'$ domains and the $\vK$ and $\vK'$-point encounters of the respective light driven trajectories are shown in red and blue, respectively. Large VP can only be achieved if excitation through the $\vK'$ point (blue trajectories) is also achieved for initial conditions from the $\vK$ domain (red area). The times $t^*$ of the Dirac point encounters for all initial conditions $\vqi$ in Fig.\,\ref{fig:pedestal}b show that this is indeed the case for the single half-cycle pulse (dashed) as well as for the one with a pedestal (solid). 
However, what difference the latter makes becomes apparent if one looks with which field strengths $F(t^*)$ these encounters happen as illustrated in Fig.\,\ref{fig:pedestal}c: 
Now it is clear that with the half-cycle pulse also unfavorable initial (blue) conditions from the $\vK'$ domain encounter the $\vK'$ point with large $F$ (dashed blue line in the blue area) and are therefore strongly excited, while for the pedestal pulse these initial conditions encounter $\vK'$ with field strength $F(t^*)\approx 0$ leaving the large field strengths encounters with high excitation to initial conditions from the $\vK$ domain. 
This implies an increase of the VP contrast and leads to the high degree of $\eta\,{=}\,72\,\%$ in Fig.\,\ref{fig:vp} as illustrated with the excitation probabilities in Fig.\,\ref{fig:pedestal}a.

To summarize, we have formulated the principles of intense-laser-pulse excitation in
gapless 2D materials. The realization that substantial excitation is only possible close to the Dirac points suggests a linearization about these points and in time leading to 
the familiar relativistic Hamilton operator for massless Dirac fermions with linear dispersion augmented by a linearized Berry connection which represents the dipole coupling to the light.
Scaling momenta and time in this effective time-dependent Hamilton operator renders it globally proportional to $\sqrt\omega$, which leads to a time-dependent Schr\"odinger equation in the scaled time $\tau = \sqrt\omega t$ completely independent of the frequency of the light, provided the pulse shape function is also formulated as a function of $\tau$.

An obvious application of this insight into the strong-field-driven dynamics of massless electrons is the control of valley polarization. As we have demonstrated, it can be achieved with half-cycle pulses linearly polarized along the vector $\vec\Delta$ connecting the $\vK$ and $\vK'$ Dirac points to break their equivalence transiently. A peak amplitude of the pulse $A_0\approx\Delta$ leads to the highest VP values. In general the VP contrast can be enhanced by increasing the wavelength to get more localized excitation in momentum space according to the scaling. A third control parameter optimizing VP is to design the pulse shape in such a way that the highest field strength during the pulse is achieved when laser-driven electron trajectories pass a Dirac point starting with initial conditions from the domain of the other pseudospin symmetry.

We expect these principles to also hold for systems with band gaps small in relation to the strong driving laser field, a conjecture which is supported by the report of VP with linearly-polarized short pulses in the gapped hBN and MoS$_2$ systems \cite{jisi+21} for ponderomotive energies considerably larger than the band gap.
Investigations of intense pulse dynamics in 2D-systems with more exotic topological properties are underway to explore the limitations and possible extensions of the universal behavior induced by the relativistic massless dynamics.

\vfill

\begin{thebibliography}{10}
\bibitem{two-color}
We use the same two-color pulse
  $\vA(t)=\frac{A_0}{\sqrt{8}}\bigs[\ve_x[\cos(\omega t)+\cos(2\omega
  t)]+\ve_y[\sin(\omega t){-}\sin(2\omega t)]\bigs]$ for
  $\lambda\,{=}\,6\,\mu$m, that has been used before \cite{mrji+21}.
\end{thebibliography}
\def\articletitle#1{\emph{#1.}}

\end{document}